
\hoffset 0.cm
\voffset0.cm
\tolerance=2000
\hbadness=2000
\vbadness=10000
\parindent=25.pt

\font\eightrm=cmr8
\font\sixrm=cmr6
\font\fiverm=cmr5
\font\eightbf=cmbx8
\font\sixbf=cmbx6
\font\fivebf=cmbx5
\font\eighti=cmmi8  \skewchar\eighti='177
\font\sixi=cmmi6    \skewchar\sixi='177
\font\fivei=cmmi5
\font\eightsy=cmsy8 \skewchar\eightsy='60
\font\sixsy=cmsy6   \skewchar\sixsy='60
\font\fivesy=cmsy5
\font\eightit=cmti8
\font\eightsl=cmsl8
\font\eighttt=cmtt8
\font\tenfrak=eufm10
\font\sevenfrak=eufm7
\font\fivefrak=eufm5
\font\tenbb=msbm10
\font\sevenbb=msbm7
\font\fivebb=msbm5


\newfam\bbfam
\textfont\bbfam=\tenbb
\scriptfont\bbfam=\sevenbb
\scriptscriptfont\bbfam=\fivebb

\newfam\frakfam
\textfont\frakfam=\tenfrak
\scriptfont\frakfam=\sevenfrak
\scriptscriptfont\frakfam=\fivefrak


\def\eightpoint{%
\textfont0=\eightrm   \scriptfont0=\sixrm
\scriptscriptfont0=\fiverm  \def\rm{\fam0\eightrm}%
\textfont1=\eighti   \scriptfont1=\sixi
\scriptscriptfont1=\fivei  \def\oldstyle{\fam1\eighti}%
\textfont2=\eightsy   \scriptfont2=\sixsy
\scriptscriptfont2=\fivesy
\textfont\itfam=\eightit  \def\it{\fam\itfam\eightit}%
\textfont\slfam=\eightsl  \def\sl{\fam\slfam\eightsl}%
\textfont\ttfam=\eighttt  \def\tt{\fam\ttfam\eighttt}%
\textfont\bffam=\eightbf   \scriptfont\bffam=\sixbf
\scriptscriptfont\bffam=\fivebf  \def\bf{\fam\bffam\eightbf}%
\abovedisplayskip=9pt plus 2pt minus 6pt
\belowdisplayskip=\abovedisplayskip
\abovedisplayshortskip=0pt plus 2pt
\belowdisplayshortskip=5pt plus2pt minus 3pt
\smallskipamount=2pt plus 1pt minus 1pt
\medskipamount=4pt plus 2pt minus 2pt
\bigskipamount=9pt plus4pt minus 4pt
\setbox\strutbox=\hbox{\vrule height 7pt depth 2pt width 0pt}%
\normalbaselineskip=9pt \normalbaselines
\rm}


\def\hcorrection#1{\advance\hoffset by #1}
\def\vcorrection#1{\advance\voffset by #1}

\newcount\notenumber  \notenumber=1              
\newif\iftitlepage   \titlepagetrue              
\newtoks\titlepagefoot     \titlepagefoot={\hfil}
\newtoks\otherpagesfoot    \otherpagesfoot={\hfil\tenrm\folio\hfil}
\footline={\iftitlepage\the\titlepagefoot\global\titlepagefalse
           \else\the\otherpagesfoot\fi}

\def\abstract#1{{\parindent=30pt\narrower\noindent\eightpoint\openup
2pt #1\par}}


\def\note#1{\unskip\footnote{$^{\the\notenumber}$}
{\eightpoint\openup 1pt
#1}\global\advance\notenumber by 1}

\def\frac#1#2{{#1\over#2}}

\def\({\left(}
\def\){\right)}
\def\<{\langle}
\def\>{\rangle}
\def\2pd#1#2#3{\frac{\partial^2#1}{\partial#2\partial#3}}

\def\sqr#1#2{{\vcenter{\vbox{\hrule height.#2pt
        \hbox{\vrule width.#2pt height#1pt \kern#1pt
           \vrule width.#2pt}
        \hrule height.#2pt}}}}


\font\bigfett=cmbx10 scaled\magstep2

\def\pslash{\not{\hbox{\kern-2.3pt $p$}}}
\def\qslash{\not{\hbox{\kern-2.0pt $q$}}}

\magnification=\magstep1
\hsize 13cm
\baselineskip=0.33in plus .5pt minus .5pt
\parskip=15.pt

\rightline{MZ-TH/92-18}
\rightline{TTP92-18}
\vskip0.3cm
\centerline {\bigfett A NOTE ON CHARGE QUANTIZATION}\bigskip
\centerline {\bigfett THROUGH ANOMALY CANCELLATION}\bigskip
 \bigskip\bigskip
\centerline {{\bf M.~Nowakowski$^{(a),}$} \note{Present address:
Physical Research Laboratory, Navrangpura, Ahmedabad 380 009, INDIA} {\it and}
{\bf A.~Pilaftsis$^{(b),}$} \note{Address after 1,
Oct.~1993,
Rutherford Appleton Laboratory, Chilton, Didcot, Oxon, ENGLAND.
E-mail address: pilaftsis@vipmza.physik.uni-mainz.de}}

\centerline {$^{(a)}$Inst.~f\"ur Theoretische Teilchenphysik, Universit\"at
Karlsruhe, 7500 Karlsruhe, {\it FRG}}

\centerline {$^{(b)}$Inst. f\"ur Physik, Johannes-Gutenberg Universit\"at,
6500 Mainz, {\it FRG}}

\bigskip
\centerline{ To appear in {\it Physical Review D}}
\bigskip
\centerline{\bf ABSTRACT} In a minimal extension of the Standard Model,
in which new neutral fermions
have been introduced, we show that the requirement of vanishing anomalies
fixes the hypercharges of all fermions uniquely.
This naturally leads to electric charge quantization in this  minimal
scenario which has features similar to the Standard Model: invariance
under the gauge group $SU(2)_L\otimes U(1)_Y$, conservation of the total
lepton number and masslessness for the ordinary neutrinos. Such minimal
models might arise as low-energy realizations of some heterotic
superstring models or $SO(10)$ grand unified theories.
\vfill\eject
\smallskip
Recently it has been shown [1-6] that the charge quantization emerges
naturally in the one-generation Standard Model (SM) and some of its extensions
from the requirement of vanishing
anomalies.  This
is quite a remarkable discovery if we recall that one of the successes
of gauge theories is their property that they are renormalizable {\it if}
all anomalies present in the theory vanish identically. Thus, in principle,
for the mechanism of charge quantization we do not impose new constraints
on the theory. However, most of the papers which discuss this subject
restrict themselves to the one generation case. Then their salient
conculsion is that the charge quantization works in SM with one generation
only if the right handed neutrino is absent i.e. the Dirac-neutrino is
massless. This might give the impression that
such considerations of one generation case tacitly assume that the
conculsion holds also for arbitrary number of generations ($N_G$). The
assumption would be that the generations are replicas of each other as
far as charge quantization is concerned. Whether this is indeed the case
is an important question in view of the Solar Neutrino Problem [7], the
Simpson Neutrino [8] or even the suggested resolution of the $\tau$-puzzle
[9] by a heavy $4th$-generation neutrino [10]. In fact,
Refs.~[5-6] which
take into account more than one generation arrive at the conclusion that
the charge is de-quantized in SM with massless Dirac-neutrinos. In [5] and
the subsequent publications [6] the proof of this statement relies on
general arguments which have to do with global, gaugeable, symmetries of the SM
Lagrangian (i.e.~the abelian lepton and baryon number
conservation on classical level).

In this note we wish to show that models which extend the SM by adding
neutral left-handed and right-handed singlets can naturally account for
charge quantization in nature. Such models could be realized in the low-energy
limit of the heterotic superstring~[11-12] or in $SO(10)$ grand unified
theories (GUT's)~[13]. It is also important to notice that such low-energy
realizations possess atractive features similar to that of the minimal SM.
i.e.,
apart from their invariance under the standard gauge group $SU(2)_L \otimes
U(1)_Y$, they conserve total lepton number and provide massless neutrinos
to any order of perturbation theory.

Next, we give a short description of the basic low-energy structure of the
model. The relevant part of the Yukawa sector containing the masses of the
leptons is given by
$$-{\cal L}^{leptons}_Y\ =\ {\phi \over v} \bar{f}_{L_i} m_{l_{ij}}
l_{R_j}\ +\ {\tilde{\phi} \over v} \bar{f}_{L_i} m_{D_{ij}} \nu_{R_j}\
+\ \bar{S}_{L_i} M_{ij} \nu_{R_j}\ +\ h.c.\eqno(1)$$
with
$$f_{L_i}\ =\ \pmatrix{ \nu_{L_i} \cr \l_{L_i} \cr} \eqno(2)$$
In Eq.~(1) we assume the presence of two singlet neutral fermions
$\nu_R$ and $S_L$ for each chiral family. The dimensional $N_G \times
N_G$ matrix $M$ can always be brought in a diagonal form by appropriate
unitary rotation of the fields $S_L$ and $\nu_R$. Similarly, the
charged lepton-mass matrix $m_l$ can also be taken to be diagonal by a
basis transformation of the fields $f_{L_i}$ and $l_{R_i}$. However,
in this certain weak basis, $m_D$ is in general a {\it nondiagonal}
$N_G\times N_G$ matrix, since there is not freedom anymore to rotate the
fields $\nu_{L_i}$ and $\nu_{R_i}$. The remaining field content for the
quarks is completely standard. The Dirac mass matrix $m_D$ is proportional
to the vacuum expectation value (VEV) $v$ of the Higgs doublet $\phi$,
while $M_{ij}$ can, for simplicity, be considered to be bare masses.
Therefore, no extra Higgs fields are, in principle, required beyond the
standard one. The neutrino mass matrix of the model mentioned above
provides the existence of $N_G$ massless neutrinos, whereas the other
$2N_G$ Weyl fermions pair up into $N_G$ heavy {\it Dirac}
neutrinos due to the total
lepton-number conservation~[12]. A surprising feature of this model
is the violation of the individual leptonic flavours in spite of the fact
that the ordinary known neutrinos are massless [12]. Since the Dirac mass
matrix $m_D$ as predicted by the model under discussion, is generally
non-diagonal, its intergenerational mixing structure can naturally explain
possible lepton-number violating and $CP$-violating signals~[14] at the
$Z^0$ peak.

In an $SU(N_C) \otimes SU(2)_L \otimes U(1)_Y$ gauge theory the conditions
for the anomalies to vanish read [15]
$$Tr\left[U(1)_Y SU(N_C)^2 \right]\ =\ \sum_{l.h.p., quarks} Y_L \ -\
\sum_{r.h.p.,quarks} Y_R\ {\buildrel ! \over =}\ 0 \eqno(3)$$
$$Tr\left[U(1)_YSU(2)_L^2\right]\ =\ \sum_{all\ doublets} Y_L \ {\buildrel
! \over =}\ 0 \eqno(4)$$
$$Tr\left[U(1)_Y^3\right]\ =\ \sum_{l.h.p.}Y_L^3\ -\ \sum_{r.h.p.}
Y_R^3\ {\buildrel ! \over =}\ 0 \eqno(5)$$
$$Tr\left[({\rm graviton})^2U(1)_Y\right]\ \propto \ \sum_{l.h.p.}Y_L\ -\
\sum_{r.h.p.}Y_R\ {\buildrel ! \over =}\ 0 \eqno(6)$$
where $l.h.p.$ ($r.h.p.$) is a shorthand notation for left handed parts
(right handed parts) of the fields assigned according to the
chirality operator. For reasons explained in [3] (large
$N_C$ expansion) we will also keep the number of colours
$N_C$ as a free parameter.
Eq.~(4) is the gravitational anomaly condition~[16] which does not play
any crucial role here. We will only occasionaly mention it.
Finally,
it should be emphasized that anomaly conditions (3)-(6) are trace
expressions and should be therefore invariant under global basis rotations in
the weak space.

In a selfexplanatory way, we introduce the hypercharge assignments
$$Y_{\phi},\ Y_{q_i}^L,\ Y_{u_i}^R,\ Y_{d_i}^R,\ Y_{f_i}^L,\ Y_{e_i}^R,
\ Y_{\nu_i}^R, \ Y_{S_i} \eqno(7)$$
for the Higgs doublet, the left handed quark doublets and the u- and
d-type quarks singlets as well as for the leptons and
the additional neutral singlets (the index $i$ runs from
$1$ to $N_G$). Then, Equations (3)--(6) can be explictly written as
$$\sum_{i=1}^{N_G}\left[2Y_{q_i}^L\ -\ Y_{u_i}^R\ -\ Y_{d_i}^R\right]\ =\ 0
\eqno(8)$$
$$\sum_{i=1}^{N_G}\left[N_C Y_{q_i}^L\ +\ Y_{f_i}^L\right]\ =\ 0 \eqno(9)$$
$$\eqalign{\sum_{i=1}^{N_G}&[2N_C (Y_{q_i}^L)^3\ -\ N_C (Y_{u_i}^R)^3\ -\ N_C
(Y_{d_i}^R)^3\cr
& +\ 2(Y_{f_i}^L)^3\ +\ (Y_{S_i})^3\ -\ (Y_{e_i}^R)^3\ -\ (Y_{\nu_i}^R)^3
]\ =\ 0 \cr} \eqno(10)$$
$$\sum_{i=1}^{N_G}\left[2N_C Y_{q_i}^L\ -\ N_C Y_{u_i}^R\ -\ N_C Y_{d_i}^R
\ +\ 2Y_{f_i}^L\ +\ Y_{S_i}\ -\ Y_{e_i}^R\ -\
Y_{\nu_i}^R\right]\ =\ 0 \eqno(11)$$
The anomaly Eqs.~(8)--(11) are not the only source of information on
hypercharges in the present model. The gauge invariance of the Yukawa terms
$$\eqalign{{\cal  L}_{Yukawa}^{quarks}\ &=\ {\cal L}_{\phi}^{quarks}\ +\
{\cal L}_{\tilde{\phi}}^{quarks} \cr
&=\ -{1 \over v}\sum_{\alpha,\beta}^{N_G}\left\{\overline{q}_{L\alpha}
\phi M^D_{\alpha
\beta}d_{R \beta}\ +\ \overline{q}_{L \alpha} \tilde{\phi} M^U_{
\alpha \beta} u_{R \beta}\right\}\ +\ h.c. \cr}\eqno(12)$$
dictates a relation between $Y_{\phi}$ and and the quark hypercharge
quantum numbers. From the first term in (12) which couples to $\phi$
it follows that $U(1)_Y$ gauge invariance is verified if
$$Y_{q_i}^L\ -\ Y_{\phi}\ -\ Y_{d_j}^R\ =\ 0 \eqno(13)$$
where $i,j=1,...N_G$ which includes $i=j$ as well as $i \not= j$. From
the second term with the dual field $\tilde{\phi}$ one gets
$$Y_{q_i}^L\ +\ Y_{\phi}\ -\ Y_{u_j}^R\ =\ 0 \eqno(14)$$
Both equations imply that
$$\eqalign{&Y_{q_1}^L\ =\ Y_{q_2}^L\ =\ Y_{q_3}^L\ =\ ...\equiv \ Y_q^L \cr
&Y_{u_1}^R\ =\ Y_{u_2}^R\ =\ Y_{u_3}^R\ =\ ...\ \equiv \ Y_u^R\ =\ Y_q^L\
+\ Y_{\phi} \cr
&Y_{d_1}^R\ =\ Y_{d_2}^R\ =\ Y_{d_3}^R\ =\ ...\ \equiv \ Y_d^R\ =\ Y_q^L\ -
\ Y_{\phi}\cr}\eqno(15)$$
Eq.~(15) is also valid when not all elements of $M^D$ and $M^U$ are
different from zero.
For example, Fritzsch-type mass matrices~[17] satisfy the above
requirement. Other viable scenarios in GUT models are mass matrices of
democratic type~[18], where all elements of $M^U$ and $M^D$ are nonzero.
This also corresponds to the fact that {\it mass-matrix elements should
generally not vanish without invoking any additional horizontal symmetry
beyond the gauge-group symmetry of the model under consideration}. In order to
{\it naturally} require some matrix elements to vanish, one has to introduce
extra Higgs fields obeying certain discrete symmetries. This extension,
however, is beyond the minimal realization for a wide class of models
considered in this work.

The situation for leptons is a little bit more involved.
Consider the Lagrangian of Eq.~(1) written in the form
$${\cal  L}_{Y}^{leptons}\ =\ {\cal L}_{\phi}^{leptons}\ +\
{\cal L}_{\tilde{\phi}}^{leptons}\ +\ {\cal L}_{Bare}^{leptons} \eqno(16)$$
Then, all the terms ${\cal L}_{\phi}^{leptons}$,
${\cal L}_{\tilde{\phi}}^{leptons}$ and ${\cal L}_{Bare}^{leptons}$
must be separately gauge invariant under the $U(1)_Y$ gauge transformations
$$\eqalign{&f_{L \alpha}\ \to \ e^{-iY_{f\alpha}^L\beta (x)} f_{L\alpha}\cr
&l_{R\alpha}\ \to \ e^{-iY_{e\alpha}^R\beta (x)} l_{R\alpha}\cr
&\nu_{R\alpha}\ \to \ e^{-iY_{\nu \alpha}^R\beta (x)} \nu_{R\alpha}
\cr
&S_{L\alpha}\ \to \ e^{-iY_{S\alpha}\beta (x)} S_{L\alpha}\cr
&\phi \ \to \ e^{-iY_{\phi}\beta (x)} \phi \cr
&\tilde{\phi}\ \to \ e^{iY_\phi \beta (x)} \tilde{\phi}\cr}\eqno(17)$$
Thus, from ${\cal L}_\phi^{leptons}$ we obtain
$$Y_{f_i}^L\ -\ Y_{\phi}\ -\ Y_{e_i}^R\ =\ 0 \eqno(18)$$
On the other hand, from ${\cal L}_{\tilde{\phi}}^{leptons}$ we get
$$Y_{f_i}^L\ +\ Y_{\phi}\ -\ Y_{\nu_j}^R\ =\ 0 \eqno(19)$$
Similar to the case of quarks, both Eqs.~(18) and (19) lead to
$$\eqalign{&Y_{f_1}^L\ =\ Y_{f_2}^L\ =\ Y_{f_3}^L\ =\ ...\equiv \ Y_f^L \cr
&Y_{e_1}^R\ =\ Y_{e_2}^R\ =\ Y_{e_3}^R\ =\ ...\ \equiv \ Y_e^R\ =\ Y_f^L\
-\ Y_{\phi} \cr
&Y_{\nu_1}^R\ =\ Y_{\nu_2}^R\ =\ Y_{\nu_3}^R\ =\ ...\ \equiv \ Y_{\nu}^R\ =\
Y_f^L\ +
\ Y_{\phi}\cr}\eqno(20)$$
Assuming the strong connection between quark-mass matrices and lepton-mass
matrices as dictated by $SO(10)$ models (i.e.~they are proportional to each
other), it is then obvious that hypercharge for each lepton family is uniquely
determined, i.e. $Y^L_f$. Finally, $U(1)_Y$~invariance of
${\cal L}_{Bare}^{leptons}$ impose the restriction
$$Y_{S_i}\ =\ Y^R_\nu\ =\ Y^L_f\ +\ Y_\phi \eqno(21)$$

It is now an easy task for us to discuss the solution of (8)-(10).
On account of Eq.~(15), condition~(8) is trivially fulfilled
since
$$\eqalign{\sum_{i=1}^{N_G}\left[2Y_{q_i}^L\ -\ Y_{u_i}^R\ -\ Y_{d_i}^R
\right]\ &=\ N_G\left[2Y_q^L\ -\ Y_u^R\ -\ Y_d^R\right]\cr
&=\ N_G\left[2Y_q^L\ -\ Y_q^L\ -\ Y_{\phi}\ -\ Y_q^L\ +\ Y_{\phi}\right]
\ =\ 0 \cr}\eqno(22)$$
whereas Eq.~(9) yields
$$\eqalign{\sum_{i=1}^{N_G}\left[N_C Y_{q_i}^L\ +\ Y_{f_i}^L \right]
\ &=\ N_G\left[N_C Y_q^L \ +\ Y_f^L \right]\ =\ 0 \cr
\Rightarrow \  Y_f^L\ &=\ -N_CY_q^L \cr}\eqno(23)$$
Condition~(10) can now be written down as follows:
$$2N_C(Y^L_q)^3-N_C(Y^R_u)^3-N_C(Y^R_d)^3+2(Y^L_f)^3+(Y_S)^3-(Y^R_e)^3-
(Y^R_\nu)^3\ =\ 0 \eqno(24)$$
By taking into account Eqs.~(15), (20) and (21), we arrive at the following
expression:
$$(Y_f^L)^3\ +\ 3Y_{\phi}(Y_f^L)^2\ +\ 3Y_{\phi}^2Y_f^L\ +\ Y_{
\phi}^3 \ =\ 0 \eqno(25)$$
which is independent of $N_C$. This relation is similar with that which has
been originally derived by Geng and Marshak for the one-generation SM~[1].
It is now straightforward to see
that Eq.~(25) has a {\it unique} solution, namely
$$Y_f^L\ =\ -Y_{\phi} \eqno(26)$$
Here, we must remark that regardless condition~(8) gravitational anomaly (6)
alone leads to the same result.
One then gets the correct value for the fermion charges by
$$Q\ =\ I_3\ +\ {1 \over \displaystyle{2 Y_{\phi}}} Y \eqno(27)$$
together with $N_C=3$ and the Eqs.~(15), (20).
We complete our proof by noting that the presence of $S_{L_i}$ in addition
to $\nu_{R_i}$ is generally necessary in such an extension of the SM.
Indeed, if all singlets $S_{L_i}$ are absent and the number of $\nu_{R_i}$
equals the number of $\nu_{L_i}$, the Lagrangian has a gaugable $B-L$
(i.e.~baryon--lepton number) symmetry and charge will be dequantized~[5,6].
If all $S_{L_i}$ and $\nu_{R_i}$ are absent, we recover the prediction for
the minimal SM. Analytically, Eq.~(25) is also valid when one considers a
general non-diagonal charged lepton-mass matrix
(e.g.~the mass matrix of democratic
type mentioned above), but {\it this is not sufficient to restore charge
quantization}, since our analysis would then be weak-basis dependent.
As mentioned above, the conditions for vanishing anomalies~(3)--(6) are trace
expressions and are invariant under global chiral rotations. This
observation allows us to show that the three-generation SM possesses hidden
gaugeable symmetries, i.e.~$L_e-L_\mu$ or $L_e-L_\tau$ or $L_\mu-L_\tau$,
and the conclusion of charge de-quantization in the SM follows~[5].
This indicates that {\it the presence of $S_{L_i}$ is absolutely necessary
to obtain charge quantization,
since these neutral singlets fix the weak space and protect it from the above
chiral rotations which cannot take place whithout affecting the $U(1)_Y$
symmetry of the model}.

For completeness we mention the case of Majorana interactions~[19,20].
Here we have to include in the Lagrangian additional mass terms of the form
$$\eqalign{&{\cal L}_{Majorana}^{\nu}\ =\ -{1 \over 2}
\bar{\nu}_{R_i} m_{M_{ij}} (\nu_{R_j})^C\ -\ {1 \over 2}\bar{S}_{L_i}
\mu_{ij} (S_{L_j})^C\ +\ h.c. \cr
&(\nu_R)^C\ =\ C(\bar{\nu}_R)^T \quad , \quad
(S_L)^C\ =\ C(\bar{S}_L)^T \cr} \eqno(28)$$
with $C$ being the charge conjugation operator. These models lead to
Majorana neutrinos even in the absence of the $m_{M_{ij}}$ term (the
latter is then sometimes called "$\mu$"-model [19]).
Of course, such models
break in general the total lepton number $L$.
Here, under $U(1)_Y$ gauge
transformations $S_{L_i}$, $\nu_{R_i}$ transform according to (17) and the
corresponding transformation for their charge conjugate fields are given by
$$\eqalign{&(\nu_{R_j})^C \ =\ e^{iY_{\nu_j}^R \beta(x)} (\nu_{R_j})^C \cr
&(S_{L_j})^C \ =\ e^{iY_{S_j} \beta(x)} (S_{L_j})^C \cr}\eqno(29)$$
Imposing $U(1)_Y$-gauge invariance on ${\cal L}_{Majorana}^{\nu}$,
one then gets $Y^R_\nu=Y_S=0$ and charge is quantized.
The phenomenology of the models considering right-handed neutrinos in the
SM has been recently investigated in~[20,21], which can account, among others,
for possible lepton-number violating effects in high energy
collider machines.

However, {\it the presence of Majorana interactions as described by
${\cal L}_{Majorana}^{\nu}$ is not a sufficient condition to restore charge
quantization}.
To give such a counter-example example let us write down a
`bare-bones-like' model similar to Ref.~[22]. Here we have
$$\eqalign{& m_D\ =\ diag(m_1,m_2,m_3) \cr
&m_l\ =\ diag(m_e,m_{\mu}, m_{\tau}) \cr}\eqno(30)$$
and
$$m_M\ =\ \left(\matrix{0&a&0 \cr a&0&b \cr 0&b&0 \cr}\right) \eqno(31)$$
For the sake of illustration, we assume that $S_{L_i}$ are absent.
{}From gauge invariance one obtains the following set of equations
$$\eqalign{&Y^R_{\nu_e}=Y^R_{\nu_{\tau}}=-Y^R_{\nu_{\mu}}\equiv Y^R
\cr
&Y^L_{f_{\tau}}=Y^L_{f_e}=Y^R-Y_{\phi},\ \ \ Y^L_{f_{\mu}}=-Y^R-
Y_{\phi} \cr
&Y^R_e=Y^R_{\tau}=Y^R-2Y_{\phi},\ \ \ Y^R_{\mu}=-Y^R-2Y_{\phi}\cr}
\eqno(32)$$
However, the requirement of anomaly cancellations yields now `only'
$$9Y_q^L+Y^R-3Y_{\phi}=0 \eqno(33)$$
Clearly $Y^R$ remains undetermined and the charge would be
de-quantized. This can be attributed to the fact that the
Yukawa sector as described by the special form of Eqs.~(30) and~(31)
possesses the local hidden symmetry $L=L_e-L_\mu+L_\tau$ which dequantizes
charge. Due to this hidden symmetry, the Weyl mass-eigenstates are
degenerate in pairs and hence form Dirac neutrino fields. Nevertheless,
if in this certain weak basis we assume that the matrix $m_M$ contains five
nonzero elements, for example, the above symmetry spoils and charge is
quantized. Furthermore, one then gets a mass spectrum of six non-degenerate
Majorana neutrinos.

In summary, there are generally
three minimal extensions of the Standard Model
by including new neutral singlets, in which the
charge of fermions is quantized:\smallskip
(i) SM with {\it non-degenerate} Majorana neutrinos.

(ii) SM with Dirac neutrinos where, however, the number of right handed
neutrino fields ($N_R^{\nu}$) must not equal the number of left handed
ones ($N_L^{\nu}$) and with the additional restriction $N_R^{\nu} \not= 0$
(see also~[6]).

(iii) the models with additional left- and right-handed neutral singlets,
where quark- and lepton-mass matrices are proportional to each other as
dictated by a wide class of GUT models. The last scenario has been
discussed in the present work.
\smallskip

\bigskip\bigskip\bigskip
{\bf Acknowledgements.} We thank J.~W.~F.~Valle  for drawing our attention
to these models and for useful discussion on this subject.
We also thank F.~Scheck, R.~Decker and
N.~Papadopoulos for valuable discussions and
comments. The
work of A.P. has been supported by a grant from the Postdoctoral
Graduate College of Mainz and the work of M.N.
by Bundesministerium f\"ur Forschung und Technologie under the Grant Nr.
06KA757.
\vfill\eject
\centerline{\bf REFERENCES}\bigskip\bigskip
[1] C.~Q.~Geng, R.~E.~Marshak, Phys.~Rev.~{\bf D39} (1989) 693;
Phys.~Rev.~{\bf D41} (1990) 717; Comments~Nucl.~Part.~Phys.~{\bf 18} (1989)
331;
C.~Q.~Geng, Phys.~Rev.~{\bf D41} (1990) 1292.\smallskip
[2] K.~S.~Babu, R.~N.~Mohapatra, Phys.~Rev.~Lett.~{\bf 63} (1989) 938;
Phys.~Rev.~{\bf D41} (1990) 271;
S.~Rudaz, Phys.~Rev.~{\bf D41} (1990) 2619;
E.~Golowich, P.~B.~Pal, Phys.~Rev.~{\bf 41} (1990) 3537;
J.~Hucks, Phys.~Rev.~{\bf D43} (1991) 2709. \smallskip
[3] A.~Abbas, J.~Phys.~{\bf G16} (1990) L163;
A.~Abbas, Phys.~Lett.~{\bf B238} (1990) 344. \smallskip
[4] N.~G.~Deshpande, University of Oregon Report No. OITS-107, 1979
(unpuplished). \smallskip
[5] R.~Foot, G.~C.~Joshi, H.~Lew, R.~R.~Volkas, Mod.~Phys.~Lett.~{\bf A5}
(1990) 95; R.~Foot, G.~C.~Joshi, H.~Lew, R.~R.~Volkas,
Mod.~Phys.~Lett.~{\bf A5} (1990) 2721.
\smallskip
[6] R.~Foot, Mod.~Phys.~Lett.~{\bf A5} (1990) 1947;
R.~Foot, Mod.~Phys.~Lett.~{\bf A6} (1991) 527;
R.~Foot, S.F.~King, Phys.~Lett~{\bf B259} (1991) 464;
J.~Sladkowski, M.~Zralek, Phys.~Rev.~{\bf D45} (1992) 1701. \smallskip
[7] L.~Wolfenstein, Phys.~Rev.~{\bf D17} (1978) 2369;
S.~P.~Mikheyev, A.~Yu~Smirnov, JETP {\bf 64} (1986) 913.
\smallskip
[8] J.~J.~Simpson, Phys.~Rev.~Lett.~{\bf 54} (1985) 1891; A.~Hime,
J.~J.~Simpson, Phys.~Rev.~{\bf D39} (1989) 1837.
\smallskip
[9] W.~J.~Marciano, Phys.~Rev.~{\bf D45} (1992), and references therein.
\smallskip
[10] M.~Shin, D.~Silverman, Phys.~Lett.~{\bf B213} (1988) 379;
S.~Rajpoot, M.~Samuel, Mod.~Phys.~Lett.~{\bf A3} (1988) 1625. \smallskip
[11] E.~Witten, Nucl.~Phys.~{\bf B268} (1986) 79;
J.~W.~F.~Valle, Proc. Intern. Symp. on Nuclear beta decays and neutrino,
Osaka, World Scientific Co., Singapore, 1986;
R.~N.~Mohapatra, J.~W.~F.~Valle, Phys.~Rev.~{\bf D34} (1986) 1642.
\smallskip
[12] J.~Bernab\`eu et al., Phys.~Lett.~{\bf B187} (1987) 303; M.~Dittmar,
J.~W.~F.~Valle, CERN yellow report CERN-91/02 (1991); M.~C.~Gonzalez-Garcia,
J.~W.~F.~ Valle, Mod.~Phys.~Lett.~{\bf A7} (1989) 385.
\smallskip
[13] D.~Wyler, L.~Wolfenstein, Nucl.~Phys.~{\bf B218} (1983) 205.
\smallskip
[14] N.~Rius, J.~W.~F.~Valle, Phys.~Lett.~{\bf B246} (1990) 249.\smallskip
[15] See for example, J.~Zinn-Justin,
`Quantum Field Theory and Critical Phenomena',
Oxford University Press, New York 1990. \smallskip
[16] E.~Witten, Phys.~Lett.~{\bf B117} (1982) 324.\smallskip
[17] H.~Fritzsch, Phys.~Lett.~{\bf B70} (1977) 436; {\bf 73} (1977) 317.
\smallskip
[18] H.~Fritzsch, Invited talk given at the International Conference
on "New Theories in Physics", Kazimierz, Poland (May 1988), MPI-PAE/Pth
60/88 (1988); H.~Harrari, H.~Haut, J.~Weyers, Phys.~Lett.~{\bf 78} (1978) 459;
Y.~Koide, Phys.~Rev.~Lett.~{\bf 47} (1981) 1241; Phys.~Rev.~{\bf D28} (1983)
252.\smallskip
[19] J.Schechter, J.~W.~F.~Valle, Phys.~Rev.~{\bf D22} (1980) 2227;
J.~W.~F. Valle, Phys.~Rev.~{\bf D27} (1983) 1672; J.~W.~F.~Valle,
Prog.~Part.~Nucl.~Phys.~ {\bf 26} (1991) 91 and references therein.
\smallskip
[20] A.~Datta, A.~Pilaftsis, Phys.~Lett.~{\bf B278} (1992) 162;
A.~Pilaftsis, Phys.~Lett.~{\bf B285} (1992) 68; Z.~Phys.~{\bf C55}
(1992) 275. \smallskip
[21] J.~G.~K\"orner, A.~Pilaftsis, K.~Schilcher, Phys.~Rev.~{\bf D47}
(1993) 1080; Phys.~Lett.~{\bf B300} (1993) 381.
\smallskip
[22] B.~Grinstein, J.~Preskill and M.~Wise, Phys.~Lett.~{\bf B159}
(1985) 57; J.~M.~Cline, Phys.~Rev.~{\bf D45} (1992) 1628.
\smallskip

\vfill\eject
\bye